\documentclass{WileyMSP-template}

\usepackage{lineno,hyperref}
\modulolinenumbers[5]

\usepackage{caption}
\usepackage{subcaption}
\usepackage{amsmath}

\begin{document}

\pagestyle{fancy}
\rhead{\includegraphics[width=2.5cm]{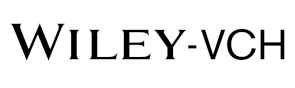}}

\title{Living Cells as a Biological Analog of Optical Tweezers \\ -- a Non-Invasive Microrheology Approach}
\maketitle

\author{William Hardiman}
\author{Matt Clark}
\author{Claire Friel}
\author{Alan Huett}
\author{Fernando P\'erez-Cota}
\author{Kerry Setchfield}
\author{Manlio Tassieri*}
\author{Amanda J Wright**}

\begin{affiliations}
W. Hardiman, M. Clark, F. P\'erez-Cota, K. Setchfield, A.J.Wright\\
Optics and Photonics Research Group, Faculty of Engineering, University of Nottingham, Nottingham NG7 2RD, UK.\\
** Email: Amanda.Wright@Nottingham.ac.uk\\

C. Friel, A. Huett\\
School of Life Sciences, University of Nottingham, Medical School, QMC, Nottingham NG7 2UH, UK.

M. Tassieri\\ 
Division of Biomedical Engineering, James Watt School of Engineering, University of Glasgow, Glasgow, G12 8LT, UK.\\
* Email: Manlio.Tassieri@Glasgow.ac.uk
\end{affiliations}


\begin{abstract}
Microrheology, the study of fluids on micron length-scales, promises to reveal insights into cellular biology, including mechanical biomarkers of disease and the interplay between biomechanics and cellular function. Here a minimally-invasive passive microrheology technique is applied to individual living cells by chemically binding a bead to the surface of a cell, and observing the mean squared displacement of the bead at timescales ranging from milliseconds to 100s of seconds. Measurements are repeated over the course of hours, and presented alongside \textit{novel} analysis to quantify changes in the cells' low-frequency elastic modulus, $G'_0$, and the cell's dynamics over the time window $\sim10^{-2}$~s to $10$~s. An analogy to optical trapping allows verification of the invariant viscosity of HeLa S3 cells under control conditions and after cytoskeletal disruption. Stiffening of the cell is observed during cytoskeletal rearrangement in the control case, and cell softening when the actin cytoskeleton is disrupted by Latrunculin B. These data correlate with conventional understanding that integrin binding and recruitment triggers cytoskeletal rearrangement. This is, to our knowledge, the first time that cell stiffening has been measured during focal adhesion maturation, and the longest time over which such stiffening has been quantified by any means.
\end{abstract}


\section{Introduction}

\noindent Single cell biomechanics is an area of increasing interest as changes in the mechanical properties of single cells have been linked to diseased and cancerous states of single cells \cite{brandao2003,Guck2005,quan2016medical}. In order to interpret bio-mechanical measurements and their implications, a better understanding is needed of the contributions of the cytoskeleton to the biomechanics of cells, and how the mechanical properties of cells are impacted by changes to the cytoskeleton. For example, targeted drugs can be used to disrupt individual cytoskeleton proteins and allowing the resultant mechanical changes to be monitored. To fully explore the interplay between single cell biomechanics and the changing properties of the cytoskeleton, novel measurement approaches are needed to observe these changes over long time courses with minimal perturbation to the living cell.

Many techniques are available to measure single cell biomechanics; some are measurements of whole cell deformation upon application of forces, whether by viscous drag\cite{brandao2003}, aspiration into micropippettes\cite{mitchison1954mechanical}, or by optical forces\cite{Guck2001}. Others, such as atomic force microscopy or Brillouin light scattering\cite{PerezCota2016,antonacci2020recent} can probe the mechanical properties of cells with sub-micron resolution. Here we chemically affix micron-sized beads to the surface of living cells to monitor the microrheology over a long time period. This not only allows study of the cell's mechanical properties on millisecond to second time-scales, but also of slow cytoskeletal dynamics on longer time-scales, all during the cell's natural interaction with a functionalised surface.

Analysis of cellular deformations requires a mechanical model of the cell, such as found in microrheology, the study of fluid flow on micron length-scales. These scales enable study of microlitre samples too small for bulk rheology and at sub-millisecond time-scales (or supra-kilohertz frequencies) inaccessible to bulk rheology due to the inertia of rheometers. Microrheology is divided into active techniques, which apply forces to the samples, and passive techniques which simply monitor thermally excited motion. The simplest passive microrheological experiment, particle tracking microrheology (PTM), is to record video of micron-sized particles suspended in the sample\cite{mason1997particle,waigh2005microrheology,waigh2016advances}. From a video of a probe particle (normally a polystyrene or silica bead), a position-time trace can be constructed and used to calculate the mean squared displacement (MSD) as a function of the time over which the displacement occurs, termed lag time, or $\tau$. Under thermal equilibrium conditions, the generalised Stokes-Einstein relation can be used to relate the MSD to the shear creep compliance \cite{mason1995optical}; and thus the MSD may be used to calculate the complex shear modulus $G^*(\omega)$.

The advantages of passive microrheology are twofold. Firstly, the full frequency spectrum of the complex modulus can be calculated from a single video recording. The upper limit to the frequencies probed is determined by the frame rate of the camera (Nyquist limit), and the lower limit may be determined by the length of the video recording or by the physics of the material probed\cite{levine2000one}. For example, living cells may undergo non-thermal strain fluctuations originating from cytoskeletal rearrangements and leading to higher MSD at long lag times (low frequencies); this limits the lowest frequencies at which passive microrheology probes mechanical properties of cells\cite{lau2003microrheology}. The second advantage is especially apparent when probing force-sensitive receptors: living cells respond to applied forces, a process known as mechanotransduction. For example, cells have been seen to stiffen under shear flow\cite{lee2006ballistic}, or temporarily soften after deformations on a short time scale\cite{lenormand2007out}. Previous studies of integrin mechanics have focused on the viscoelastic properties of the integrin binding without consideration of either how the applied forces affect the mechanics of the cell, or of how the cell's mechanics may change over many minutes\cite{bausch2001rapid,matthews2004mechanical}. Here, our passive approach allows us to probe the mechanics of the binding site as it evolves over an hour or more without perturbation by non-thermal forces.

In this work, we demonstrate non-invasive passive microrheology of living cells to probe both thermal strain fluctuations at millisecond to second timescales and cytoskeletal remodelling over longer times, without mechanically disrupting the cell. Previous methods probing cellular mechanics have either been invasive\cite{lau2003microrheology,lee2006ballistic}, did not probe mechanics at such short time-scales\cite{bursac2005cytoskeletal,bursac2007cytoskeleton,an2004role}, or relied on application of external forces\cite{bausch1998local,Fabry2001}. 

The experimental method, in short, is to chemically affix a functionalised microsphere to the surface of a living cell; the pseudo-Brownian motion of the microsphere is recorded and interpreted, with repeat measurements of the same cell over multiple hours. The method is based on the prior work of Warren et al.\cite{Warren2013} who first presented the experimental technique and analytical model. By making analogy to optical trapping, we develop a framework for quantifying changes to cell microrheology. This novel approach allows us to quantify the viscosity of the cell and to measure cytoskeletal rearrangement, not just as a strain rate, but as an equivalent to internal stress generation. Importantly this allows separation of the change in cytoskeletal activity from the effects of the cell stiffening or softening.

We first describe our analytical framework, developing an analogy between our cell measurements and microrheology with optical tweezers. Then the data collection and novel analysis is explained, followed by results from cell measurements with and without drug treatment to disrupt the actin cytoskeleton.
 
 \section{Analytical framework}\label{sec:Analytical}
\subsection{Microrheology with Optical Tweezers}

\noindent
When a micron sized spherical particle is suspended in a complex fluid and it is optically trapped (OT) by a highly focused laser beam, the thermal fluctuation of the molecules in the fluid drive the constrained Brownian motion of the particle. Its dynamics can be described analytically by means of the following generalised Langevin equation \cite{TASSIERI201939}:
\begin{equation}
\label{eq:Langevin1}
	m \vec{a}(t)=\vec{f}_{R}(t)-\int^{t}_{0}\zeta(t-\tau){\vec{v}}(\tau)d\tau -\kappa\vec{r}(t),
\end{equation}
where $m$ is the mass of the particle, $\vec{a}(t)$ is its acceleration, $\vec{v}(t)$ is its velocity, $\vec{r}(t)$ is its position relative to the trap center ($\vec{r}_c\equiv 0$), $\kappa$ is the trap stiffness and $\vec{f}_{R}(t)$ is a Gaussian white noise term, modelling stochastic thermal forces acting on the particle. The integral term, which incorporates a generalised time-dependent memory function $\zeta(t)$, represents the viscoelastic force exerted by the fluid on the particle.
Following the approach used by Mason~\&~Weitz \cite{mason1995optical} for the case of freely diffusing particles, equation~(\ref{eq:Langevin1}) can be solved for the shear complex modulus ($G^*(\omega)$) of the material in terms of either the normalised mean squared displacement (NMSD) $\Pi(\tau)=\langle\Delta r^2(\tau)\rangle / 2\left\langle r^2\right\rangle$ \cite{ISI:000275053800036} or the normalised position autocorrelation function (NPAF) $A(\tau)=\left\langle \vec{r}(t)\vec{r}(t+\tau)\right\rangle / \left\langle r^2\right\rangle$ \cite{ISI:000291926500023}:
\begin{equation}
\label{eq:G*OT}
	G^*(\omega)\frac{6\pi a}{\kappa}=\left(\frac{1}{i\omega \hat{\Pi}(\omega)}-1\right)\equiv \left(\frac{1}{i\omega\hat{A}(\omega)}-1\right)^{-1}\equiv \frac{\hat{A}(\omega)}{\hat{\Pi}(\omega)}
\end{equation}
where $\left\langle r^2\right\rangle$ is the variance of the particle's trajectory, $\hat{\Pi}(\omega)$ and $\hat{A}(\omega)$ are the Fourier transforms of $\Pi(\tau)$ and $A(\tau)$, respectively. The inertial term ($m\omega^{2}$) present in the original works has been neglected here, because for micron-sized particles it only becomes significant at frequencies of the order of MHz.
The frequency-dependent shear complex modulus of a material is a complex number $G^*(\omega)=G'(\omega)+iG''(\omega)$, whose real and imaginary parts provide information on the elastic and the viscous nature of the material under investigation~\cite{Ferry:1980jo}. These are commonly indicated as the storage ($G'(\omega)$) and the loss ($G''(\omega)$) moduli, respectively.

\subsubsection{Optical Tweezers as an ideal viscoelastic material}

At this point we shall write some of the fundamental relationships between the most common parameters describing the materials' linear viscoelastic (LVE) properties and the time-averaged functions (e.g., the MSD) derived from the analysis of the particle's thermal fluctuations.

Let us begin by describing a simple relationship between the MSD of a freely diffusing particle and the time-dependent compliance $J(t)$ of the suspending fluid.
In classical rheology (i.e.~in shear flow) the creep compliance is defined as the ratio of the time-dependent shear strain $\gamma(t)$ to the magnitude $\sigma_0$ of the constant shear stress that is switched on at time $t=0$: $J(t)=\gamma(t)/\sigma_0$. The compliance is related to the shear relaxation modulus $G(t)$ by a convolution integral~\cite{Ferry:1980jo}:
\begin{equation}
\label{convolution}
	\int_0^t G(\tau)\,J(t-\tau)\,d \tau=t.
\end{equation}
The complex shear modulus $G^*(\omega)$ is also defined as the Fourier transform of the time derivative of $G(t)$, hence by taking the Fourier transform of eq.~(\ref{convolution}) it follows that:
\begin{equation}
\label{G*J*}
	G^*(\omega) = i\omega \hat{G}(\omega)= \frac{1}{i\omega \hat{J}(\omega)}
\end{equation}
where $\hat{G}(\omega)$ and $\hat{J}(\omega)$ are the Fourier transforms of $G(t)$ and $J(t)$, respectively.
By equating equation~(\ref{G*J*}) with the following generalized Stokes-Einstein equation \cite{mason1995optical}:
\begin{equation}
\label{G*MSD}
	G^*(\omega) =\frac{k_B T}{i\omega\pi a \left\langle\Delta \widehat{r^2}(\omega)\right\rangle}
\end{equation}
one obtains:
\begin{equation}
\label{J*MSD}
	\left\langle\Delta \widehat{r^2}(\omega)\right\rangle= \frac{k_{B}T}{\pi a}\hat{J}(\omega)~~~\Longleftrightarrow~~~\left\langle\Delta r^2(\tau)\right\rangle= \frac{k_{B}T}{\pi a}J(t)
\end{equation}
where $\left\langle\Delta \widehat{r^2}(\omega)\right\rangle$ is the Fourier transform of the mean square displacement $\left\langle\Delta r^2(\tau)\right\rangle \equiv \left\langle\left[\vec{r}(t+\tau)-\vec{r}(t)\right]^{2}\right\rangle$. 
The average $\left\langle \ldots \right\rangle$ is taken over all initial times $t$ and all particles, if more than one is observed.
In equations~(\ref{G*MSD}) and (\ref{J*MSD}) it has been assumed that the inertial term ($m\omega^{2}$) is negligible for frequencies $\ll$ MHz and that $J(0)=0$ for viscoelastic fluids (i.e. no prestress).
Equation~(\ref{J*MSD}) expresses the linear relationship between the MSD of suspended spherical particles and the macroscopic creep compliance of the suspending fluid~\cite{xu1998compliance}.

In the case of an optically trapped spherical particle suspended in a generic viscoelastic fluid, equation~(\ref{J*MSD}) would still hold, but it would describe the relationship between the measured MSD of a constrained particle and the compliance ($J_{tot}$) of the compound system made up of the optical trap and the viscoelastic fluid \cite{tassieri2012microrheology}: 
\begin{equation}
\label{J*PI}
	\left\langle\Delta r^2(\tau)\right\rangle= \frac{k_{B}T}{\pi a}J_{tot}(t)~~~or~~~\Pi(\tau)=\frac{\kappa}{6\pi a}J_{tot}(t)
\end{equation}
where the second expression has been obtained by dividing the first one by twice the variance and applying the principle of equipartition of energy.
Notably, in the simplest case of a symmetric optical trap where $\kappa \equiv \kappa_j \forall j \in \{x,y,z\}$ and a suspending Newtonian fluid with a time-independent viscosity $\eta$, the compound system (OT plus fluid) can be modelled as an \emph{ideal} Kelvin-Voigt material (an elastic element in parallel with a viscous element), with elastic constant proportional to the trap stiffness, 
$\kappa / (6\pi a)$, and viscosity equal to $\eta$. In this case, $J_{tot}$ assumes a simple analytical form:
\begin{equation}
\label{KV}
	J_{tot}=\frac{6\pi a}{\kappa}\left(1- e^{-\lambda t}\right)~~~\Rightarrow~~~\Pi(\tau)=\left(1- e^{-\lambda \tau}\right)
\end{equation}
where $\lambda=\kappa / (6\pi a \eta)$ is the relaxation rate of the compound system, known as the ``\textit{corner frequency}", or $f_c$, when the thermal fluctuations of an optically trapped particle are analysed in terms of the power spectral density~\cite{berg2004power}.
In practice, $\lambda$ defines a characteristic time ($t^*=\lambda^{-1}$) at which the fluid compliance ($J(t)=t/\eta (t)$) equals the compliance of the optical trap ($J_{OT}=6\pi a/\kappa$): $J(t^*)=J_{OT}$.

Finally, from equations~(\ref{G*J*}) and~(\ref{J*PI}), one can also express the viscoelastic properties of the compound system in the frequency domain:
\begin{equation}
\label{G*PI*}
	G^{*}_{tot}(\omega)=\frac{\kappa}{6\pi a} \frac{1}{i\omega \hat{\Pi}(\omega)}.
\end{equation}
For a system modelled as an \emph{ideal} Kelvin-Voigt material the above equation becomes:
\begin{equation}
\label{G*KV}
	G^{*}_{tot}(\omega)=\kappa / (6\pi a) +i\eta \omega,
\end{equation}
for which the elastic and viscous behaviours of the system are clearly defined.
Indeed, from equation~(\ref{G*KV}) it can be seen that, (i) for $\omega \rightarrow 0$, the system behaves as an elastic solid with a zero frequency shear elastic modulus equal to: $G^{*}_{tot}(0)=\kappa / (6\pi a)$; whereas, (ii) for $\omega >> \lambda$, the viscous component governs the system's dynamics and $G^{*}_{tot}(\omega)\simeq i\eta\omega$. This behaviour can be seen by inspection of figure \ref{fig:cell_w_bead} e and f) which demonstrate the MSD and complex modulus, respectively, of an optically trapped bead in water.

The concepts introduced in this section will be built upon in the following section, which will explain how pseudo-Brownian motion can be used to observe the cells' dynamics by analysing the constraining force exerted by a cell onto a bead attached to its membrane. In particular, equation \ref{G*PI*} will be seen to have an analog in the cell-bead system.

\subsection{Cells as a biological analog of Optical Tweezers}
\noindent
Warren et al. \cite{Warren2013} adopted the theoretical framework introduced in section 2.1 to interpret the \textit{pseudo} Brownian motion of a bead chemically attached onto the surface of a cell and no longer optically trapped. In this case, the cell acts as a biological analog of Optical Tweezers as will be seen after interpretation of the MSD. First, a possible generalised Langevin equation could be written as:
\begin{equation}
    m\vec{a}(t) = \vec{f}_R(t) -  \int_0^t [\zeta_c(t - \tau) + \zeta_s(t - \tau)]\vec{v}(\tau)d\tau,
    \label{eq:LangvinWarren}
\end{equation}
where the integral term represents the total damping force
acting on the bead. Based on the superposition principle, the damping incorporates two generalised time-dependent memory
functions $\zeta_c(t)$ and $\zeta_s(t)$ that are representative of the viscoelastic nature of the cell and the solvent, respectively.
In particular the above two memory functions are related to the complex modulus of the solvent and the cell by means of the following two expressions: $G^*_s(\omega)\cong i\omega\hat{\zeta}_s(\omega)/6\pi a$ and $G^*_c(\omega)=i\omega\hat{\zeta}_c(\omega)/\beta$, respectively; where $\hat{\zeta}_s(\omega)$ and $\hat{\zeta}_c(\omega)$ are the Fourier transforms of $\zeta_s(t)$ and $\zeta_c(t)$, and $\beta$ is a constant of proportionality having dimension of a length introduced by Warren et al~\cite{Warren2013}. $\beta$ may vary for different cells as it depends on (i) the cell radius, (ii) the number and the dynamics of the chemical bonds between the bead and the cell, (iii) the contact area between the cell and the glass coverslip, and (iv) the relative position of the bead with respect to both the cell’s equatorial plane and the glass coverslip. 

By assuming the system to be at thermodynamic equilibrium, equation~(\ref{eq:LangvinWarren}) can be solved for the viscoelastic modulus of the cell in terms of the Fourier transform of the bead NMSD~\cite{Warren2013}:
\begin{equation}
     \frac{G_c^*(\omega)}{G'_0} = \frac{1}{i\omega\hat{\Pi}(\omega)}+\frac{m\omega^2}{\beta G'_0}-\frac{6\pi RG_s^*(\omega)}{\beta G'_0},
    \label{eq:LangvinWarrenSol}
\end{equation}
where \textbf{$\beta G'_0=k_BT/\left\langle r^2\right\rangle$}, is the limiting value, for vanishingly low frequencies, of the elastic modulus of the compound system (i.e. cell plus solvent), which in this work is $G'_0\equiv G_c^*(\omega)$ for $\omega \rightarrow 0$.
We make two assumptions to simplify equation~(\ref{eq:LangvinWarrenSol}): (i) for micron-sized silica beads, the inertia term $m\omega^2$ is negligible up to frequencies on the order of MHz and (ii) for solvents having frequency-independent viscosity $\eta_s$ (e.g., water), $G_s^*(\omega)$ simplifies to $i\omega\eta_s$ and the last term in equation~(\ref{eq:LangvinWarrenSol}) becomes negligible for the range of frequencies explored in this work (even with Faxén's correction to the apparent viscosity due to the proximity to the glass surface).
Thus, equation~(\ref{eq:LangvinWarrenSol}) can be further simplified into:
\begin{equation}
     \frac{G_c^*(\omega)}{G'_0} = \frac{1}{i\omega\hat{\Pi}(\omega)},
    \label{eq:LangvinWarrenSol2}
\end{equation}
which provides a means of measuring the viscoelastic properties of the cell (scaled by $G'_0$) over a range of frequencies. In practice the frequency range is limited at high frequencies by the acquisition rate of the detector used for tracking the bead position, and at low frequencies by 
cytoskeletal reorganisations driving bead motion of greater magnitudes than the thermal motion.
The equivalency between equation~(\ref{eq:LangvinWarrenSol2}) and equation~(\ref{G*PI*}) (for an optically trapped particle suspended into a generic complex fluid) provides the analogy with optical trapping previously mentioned.

\section{Microrheology experiments}

\subsection{Data collection}

\begin{figure}
    \centering
    \includegraphics[width=0.7\textwidth]{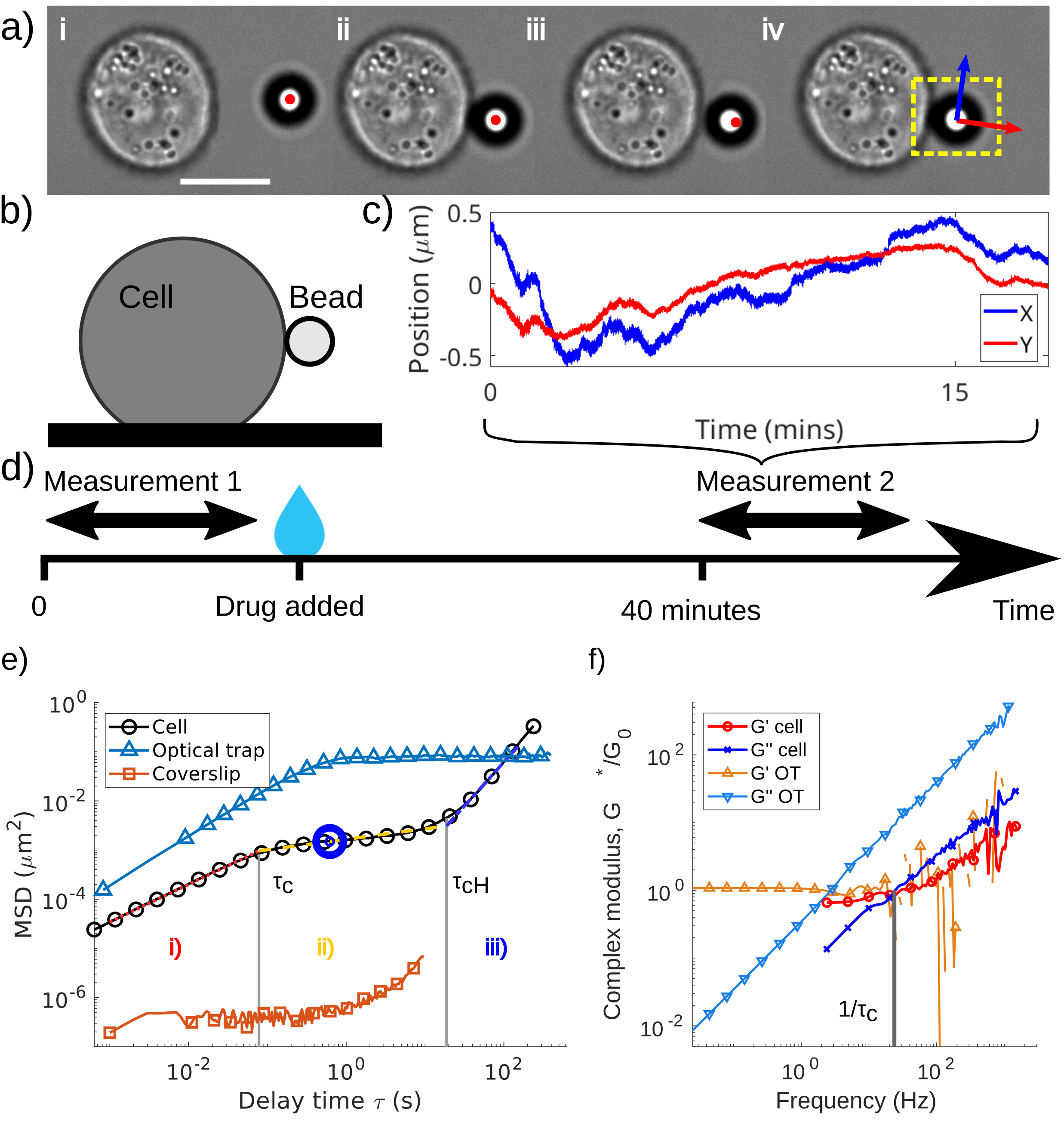}
    \caption{Data collection and interpretation. a) Experimental procedure. i. A functionalised bead is brought to the equatorial plane of the cell using the optical trap (red dot). ii. The bead is held in contact with the cell for up to 2 minutes to facilitate binding. iii. The binding is tested by pulling the bead away from the cell. iv. The optical trap is switched off and a video is recorded with a small region of interest (ROI, yellow rectangle) to increase acquisition speed. Note the arrows showing radial and tangential directions, used after a co-ordinate transform. b) Schematic side view of bead attached to cell. c) Position-time trace from 1 measurement; the 18 minute trace consists of 2,000,000 observations. d) Experimental paradigm for change over time: two video measurements are taken 40 minutes apart with a drug being added after the first measurement. e) log-log plot with three example MSD curves: bead attached to cell, optically trapped bead, and bead attached to coverslip (to demonstrate noise floor). Three regions are highlighted for the cell MSD: i) viscoelastic response at short time, ii) soft glassy plateau at intermediate time with power-law exponent minima indicated by blue circle, and iii) superdiffusion at long times. f) Normalised complex modulus of a cell (equation \ref{eq:LangvinWarrenSol2}), and of optical tweezers (equation \ref{G*PI*} divided by $\kappa$), calculated from the Fourier transform of the normalised MSD. Characteristic time scales found empirically are labelled on e) and f).}
    \label{fig:cell_w_bead} 
\end{figure}

\noindent Polystyrene beads of 5$\mu$m diameter, functionalised with streptavidin (Spherotech, USA) to facilitate binding to the cell, are added to the media once a cell is chosen for imaging. The beads take a couple of minutes to sediment to the bottom of the sample holder, then one is optically trapped (figure \ref{fig:cell_w_bead}a i)). The trapped bead is brought to the edge of the cell and the focus control on the microscope is used to adjust the position in \textit{z} relative to the cell with the ideal position shown in figure \ref{fig:cell_w_bead}b). The cell is then brought into contact with the bead (figure \ref{fig:cell_w_bead}a ii)), judged by observing the bead displacing from the centre of the trap on contact with the cell. If the bead moves in \textit{z} as contact is made then the focal position was incorrect, and the cell is moved away to adjust the \textit{z} position of the bead. Motion of the bead in \textit{z} is estimated by a change in the appearance (or brightness profile) of the bead.

Once the trapped bead is in contact with the cell, the laser power is reduced to form a weak trap ($\kappa \sim 2$ \textmu N/pm), which is used to attempt to remove the bead from the cell (figure \ref{fig:cell_w_bead}a iii)). The trap must be sufficiently weak to avoid pulling a membrane tether\cite{zhang2008optical}. If the weak trap is unable to separate the bead from the cell, binding is confirmed, the trapping laser is turned off, and a small region of interest containing just the bead is imaged, as shown in figure \ref{fig:cell_w_bead}a iv). Videos recording the motion of the bead, typically comprising $1$ million frames, are analysed by thresholding and centroiding to produce position-time tracks of the beads, an example of which can be seen in figure 1c.

During experiments, measurements were taken at regular intervals and, in order to better discriminate mechanical changes in the cell over time, two of these measurements roughly $40$ minutes apart were chosen to be analysed (figure \ref{fig:cell_w_bead}d). Drug treated cells had Latrunculin B added to the cell media dissolved in $200$ $\mu$L of DMEM at $2$mM (the same volume as the media already in the microslide chamber), thus ensuring homogeneity of the drug concentration within the sample.

\subsection{Data analysis}

\begin{figure}
    \centering
    \includegraphics[width=0.7\textwidth]{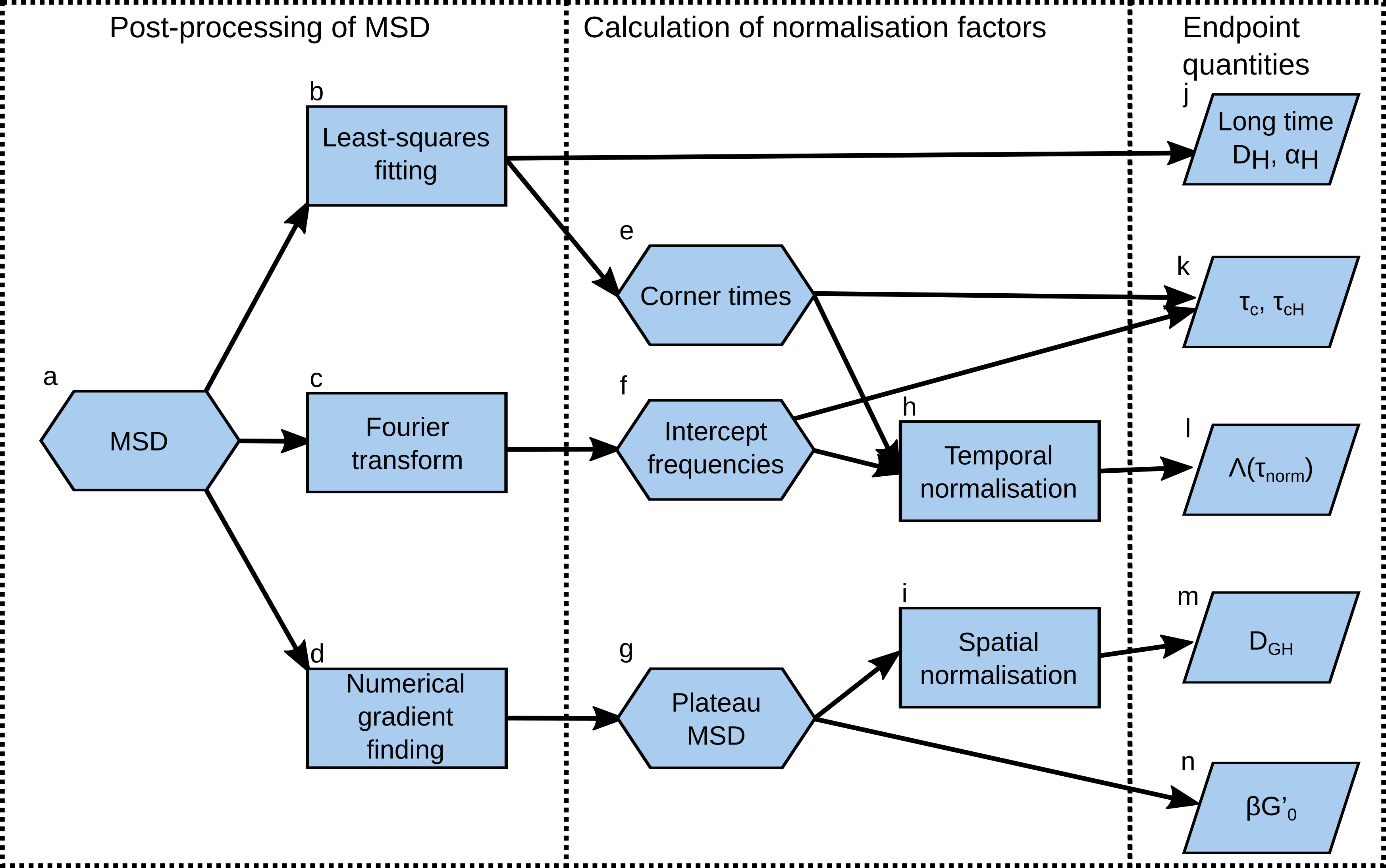}
    \caption{Post-processing steps applied to experimental data, from the MSD on the left resulting in calculation of 9 parameters listed on the right. Hexagons indicate values used during processing, rectangles indicate processing steps, rhomboids indicate the endpoint measures, and arrows indicate flow of data between the steps. Each processing step and the interpretation of the parameters is explained in detail in the main text.}
    \label{fig:flowchart}
\end{figure}

\noindent Offline data analysis was performed using MATLAB (Mathworks, USA). The videos were processed to extract the $(x,y)$ coordinates of the bead position within each frame. The Cartesian coordinate were then transformed into polar coordinates $(r,\theta)$ to separate the motion components into the radial and tangential directions, which are perpendicular and tangential to the cell surface, respectively. 
Notice that the tangential coordinate used for the analysis is defined as $\vec r\cdot\vec\theta$; therefore, both coordinates have dimensions of a length.

The MSD (see example in figure \ref{fig:cell_w_bead}e)) has been calculated by using a code adapted from the one presented by Tarantino \cite{tarantino2014tnf}. In particular, the original MATLAB code was modified to calculate the MSD not at all possible delay times (i.e., at lag times linearly spaced in time), but at lag-times logarithmically spaced in time, this significantly reduce the overall processing time. The MSD was then analysed as schematically shown in figure \ref{fig:flowchart}.

In order to evaluate the Fourier transform of experimental raw data (figure \ref{fig:flowchart} box c), an analytical methods was implemented in MATLAB based on the method presented by Evans \textit{et al}.\cite{evans2009direct}. Numerical errors at high frequencies were reduced by following Tassieri and Smith’s approach of virtually oversampling the MSD data using a cubic spline interpolation function before applying Evans’ analytical formula\cite{tassieri2012microrheology,smith2021rheoft}. 

\subsubsection{Stiffness and viscosity}

Now we explain interpretation/analysis of our experimental MSD curves,
starting from the low-frequency limit the elastic modulus of the cell, $\beta G'_0$, which is equivalent to the trap stiffness for an optically trapped bead. For an optically trapped bead this stiffness can be calibrated using the principle of equipartition of energy for which only the variance of the bead position is needed~\cite{TASSIERI201939}. Note that for optical tweezers at long lag-times, the variance of the bead position is equal to half the plateau value of the MSD~\cite{tassieri2010measuring}.
Similarly, when a bead is bound to the surface of a cell, the bead's pseudo-diffusion is constrained by the binding on timescales of the order of a few seconds. Cells, however, do not remain stationary, and the dynamics of the attached bead are governed by the cytoskeletal reorganisation over timescales longer than a second, especially when beads are bound to integrin receptors of living cells~\cite{lau2003microrheology,bursac2005cytoskeletal,bursac2007cytoskeleton}. As such, the variance ($\left\langle r^2\right\rangle$) of the bead position does not reflect the low-frequency limit of the cells' elastic modulus (as for the OT). Therefore, we have considered instead the half value of the plateau of the mean square displacement curve as a means of measuring the elastic plateau modulus of cells:
\begin{equation}
    \beta G'_0 = \frac{ 2 k_B T }{MSD_{p}},
    \label{eq:cell_stiff}
\end{equation}
where $MSD_{p}$ is the plateau value of the MSD. To find the MSD plateau (figure \ref{fig:flowchart} boxes d and g, shown in figure \ref{fig:cell_w_bead}e circled in blue), the power law exponent was estimated by means of a rolling least-squares fit 
and the plateau was taken to be coincident with the power-law exponent minimum. 

Another relevant parameter to be considered is the onset time of the MSD plateau, which in the case of an OT is given as $t^*=(6 \pi r \eta_s)/\kappa$, where $\kappa = {2 k_B T}/{MSD_p}$, and $\eta_s$ is the fluid dynamic viscosity. Similarly, for a bead bound to a cell, we propose the following expression to evaluate the cell's dynamic viscosity:
\begin{equation}
    \eta_c = \frac{\beta G'_0 \tau_c}{6 \pi r}\equiv\eta_r\eta_s,
    \label{eq:trap_corner}
\end{equation}
which can be determined by using the geometric stiffness, $\beta G'_0$, as defined in equation \ref{eq:cell_stiff}, and the corner time $\tau_c$, measured experimentally. In equation \ref{eq:trap_corner} $\eta_r=\eta_c/\eta_s$ is the relative viscosity of cells.

In order to find the corner times, $\tau_c$ and $\tau_{cH}$, from the MSD data, two methods were adopted using either the time or frequency domain and a normalisation was performed by using the mean value of these times.
In the time domain (figure \ref{fig:flowchart} boxes b and e, labelled on figure \ref{fig:cell_w_bead}e)), the corner times are estimated by the abscissa value of the intercept between the least-squares linear fits of the MSD data (drawn in a log-log plot) within the linear regions either defined by the corner times. In the frequency domain (figure \ref{fig:flowchart} boxes c and f, shown in figure \ref{fig:cell_w_bead}f as $1/\tau_c$), the corner time is estimated by the inverse of the frequency for which $tan(\delta) = G''/G' = 1$; this is estimated by means of a least-squares fit of $log(\omega)$ against $log(tan(\delta))$. 


\subsubsection{Spatio-temporal normalisation}

The identification of different characteristic timescales relevant to the cell-bead system allows time-domain normalisation. This has previously been demonstrated for optical trapping microrheology measurements\cite{tassieri2012microrheology}, where it can remove variation due to different trap stiffness or sample viscous response.

In the case of live cell microrheology, different processes drive the motion at different timescales, and the characteristic timescales vary for each cell-bead pair. Using these normalisations allows us to pool data from different cells or to remove the effect changing one parameter has on another; e.g., change in cytoskeletal reorganisation independent of stiffening of cell. For both applications, the corner times and plateau MSDs are found empirically as described above (figure \ref{fig:flowchart} boxes e, f, and g). The spatial normalisation is performed by dividing the MSD values by the plateau MSD, and the temporal normalisation by dividing the lag times by corner times. This allows better understanding of cytoskeletal reorganisation and increment distributions.

\subsubsection{Cytoskeletal reorganisation}

\noindent The long-time superdiffuse motion of the attached bead is quantified by means of least-squares linear fits of a portion of the experimental MSD data to
\begin{equation}
    log(MSD) = log(2D_H) + \alpha_H log(\tau),
    \label{eq:leastSq}
\end{equation}
providing $D_H$, the pseudo-diffusion coefficient, and $\alpha_H$, the power-law exponent, as represented by figure \ref{fig:flowchart}, box j. 


The long-time super-diffusive motion (labelled as ``iii'' in figure \ref{fig:cell_w_bead} e)) is brought about by active strain fluctuations within the cell due to cytoskeletal reorganisation. $D_H$ can be interpreted equivalently to rate of strain, or the square of the distance that the bead binding site moves in a given time. At long lag-times, $1 < \alpha_H < 2$, ruling out both thermal motion (which has a power-law exponent $\leq 1$) and experimental drift (which has power-law exponent $=2$) as the source of the superdiffuse motion. 

\subsubsection{Increment distributions}
\noindent
In order to better understand the driving forces behind bead motion, we have examined the normalised increment distribution as a function of lag time,
\begin{align}
    z(\tau) &= \left.  \frac{\Delta r - \overline{\Delta r}}{STD(\Delta r)} \right|_\tau,
\end{align}
where $\Delta r(\tau)=(r(t+\tau) - r(t))$ is the increment series for a given lag-time ($\tau$), $\overline{\Delta r}$ is its mean and $STD(\Delta r_\tau)$ is its standard deviation.
Notice that, for Brownian motion, the increments are normally distributed across all time scales, as corroborated by our measurements performed with optical tweezers and shown later in figure \ref{fig:ddist_and_ngp}. This prediction can be validated by considering the bead as an overdamped oscillator (equation \ref{eq:Langevin1}), in which the mean driving force is proportional to the displacement over a given time interval. If molecular motors have a significant driving contribution to motion, it is expected that the increment distribution will instead have a broad-tailed distribution\cite{bursac2005cytoskeletal}.

To pool data from different cells, we first calculated the increment distribution for each cell as a function of lag-time, then time-domain normalisation was performed (figure \ref{fig:flowchart} boxes h and l), and the increment distributions were added together before further analysis. The increment distributions were quantified using a non-Gaussian parameter\cite{weeks2002properties,bursac2005cytoskeletal},
\begin{equation}
    \Lambda(\tau) = \frac{\langle z(\tau)^4 \rangle}{3 \langle z(\tau)^2 \rangle ^2} - 1, \label{eq:ddist}
\end{equation}
where $z(\tau)$ is the normalised increment distribution defined above. This approach quantifies the deviation from a normal distribution, with values $\Lambda(\tau) > 0$ indicating a broad-tailed distribution.

\section{Results}

\begin{figure}
    \centering
    \includegraphics[width=0.95\textwidth]{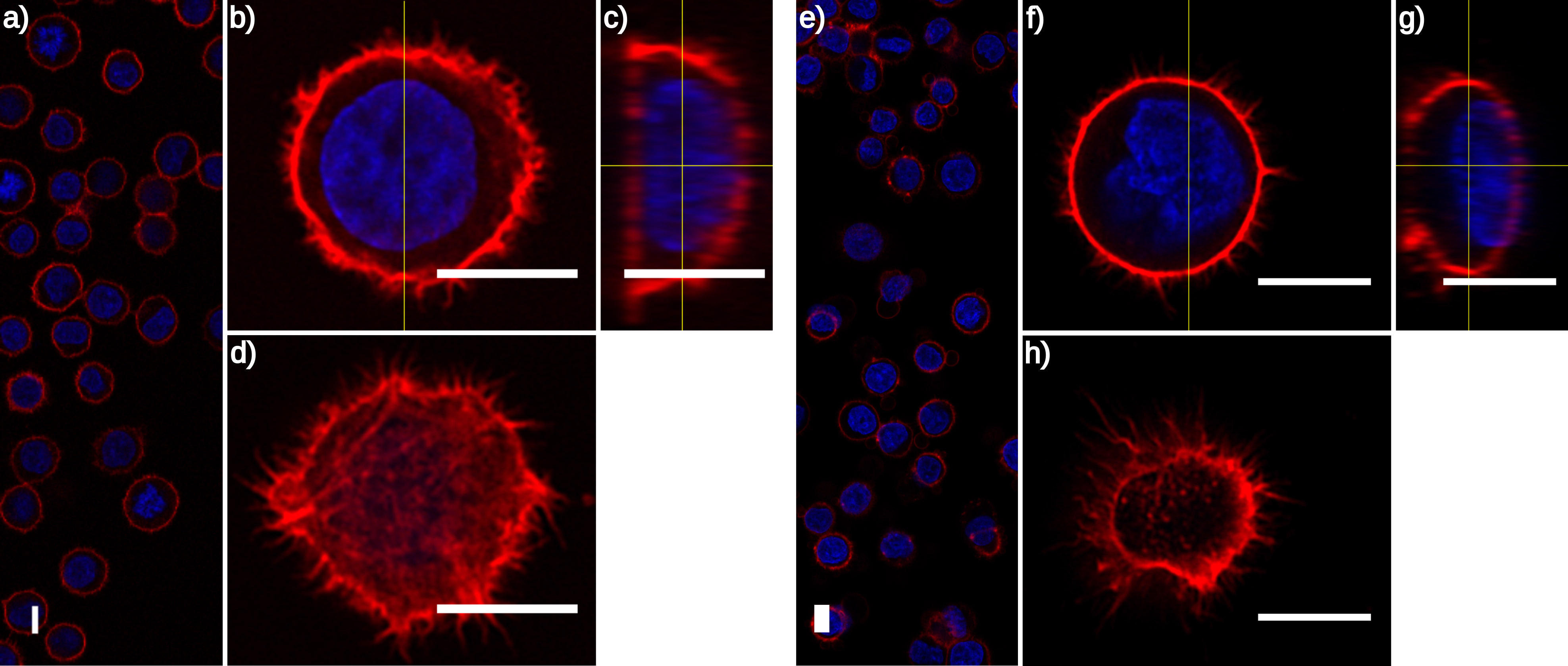}
    \caption{
    Airy scan confocal images of actin cytoskeleton (red, phalloidin) and nucleus (blue, Hoescht 33342) in fixed HeLa S3 cells (a-d) and 10 minutes after Latrunculin B treatment (e-h). (a, e) wide field of view, (b, f) medial plane with indicator of location of profile slice, (c, g) profile slice with indicator to show height of medial plane, and (d, h) basal plane. All scale bars are 10\textmu m.
    }
    \label{fig:confocal}
\end{figure}

\noindent 
First we consider the effect of Latrunculin B on Hela S3 cells visualised by means of confocal microscopy, as shown in figure \ref{fig:confocal}. Latrunculin B is known to disrupt the actin filaments in the cytoskeleton. Several changes to cell morphology are visible after drug treatment: there are blebs present on many of the cells visible in figure \ref{fig:confocal}e), the actin cortex is thinner and has fewer fibrous protrusions, the cell is less spread with less contact to the coverslip, and there is lower actin density where the cell contacts the substrate (figure \ref{fig:confocal}d vs h). Also, after drug treatment, the nucleus appears to be further from the coverslip (figure \ref{fig:confocal}c vs g), which may be caused by the reduced tension within the actin cytoskeleton.

Microrheology experiments were performed on control and drug treated Hela S3 cells and the time-dependent behaviour of the MSD curves was inspected. For measurements where $\left.MSD\right|_{\tau=1s} < 10^{-4}\mu$m$^2$, data was deemed to be too close to the static noise floor (as shown by bead attached to coverslip in figure \ref{fig:cell_w_bead}) and was discarded. From looking at transmission images recorded of the cell-bead pair it is clear that in this discarded data set the beads had been partially engulfed by the cells. After purging these measurements, there remained data from $21$ cells, $10$ of which received drug treatment.

The MSD curves were interpreted and endpoint measures were calculated (figure \ref{fig:flowchart}) for two video recordings $40$ minutes apart; when ``change over $40$ minutes'' is used herein, it refers to the difference between the values at these two time stamps. Results are compared for drug treated and control conditions, and a Mann-Whitney u-test is used to determine if the medians are significantly different.

\subsection{The viscosity and elasticity of the cells}

\begin{figure}[t]
    \centering
    \includegraphics[width=\columnwidth]{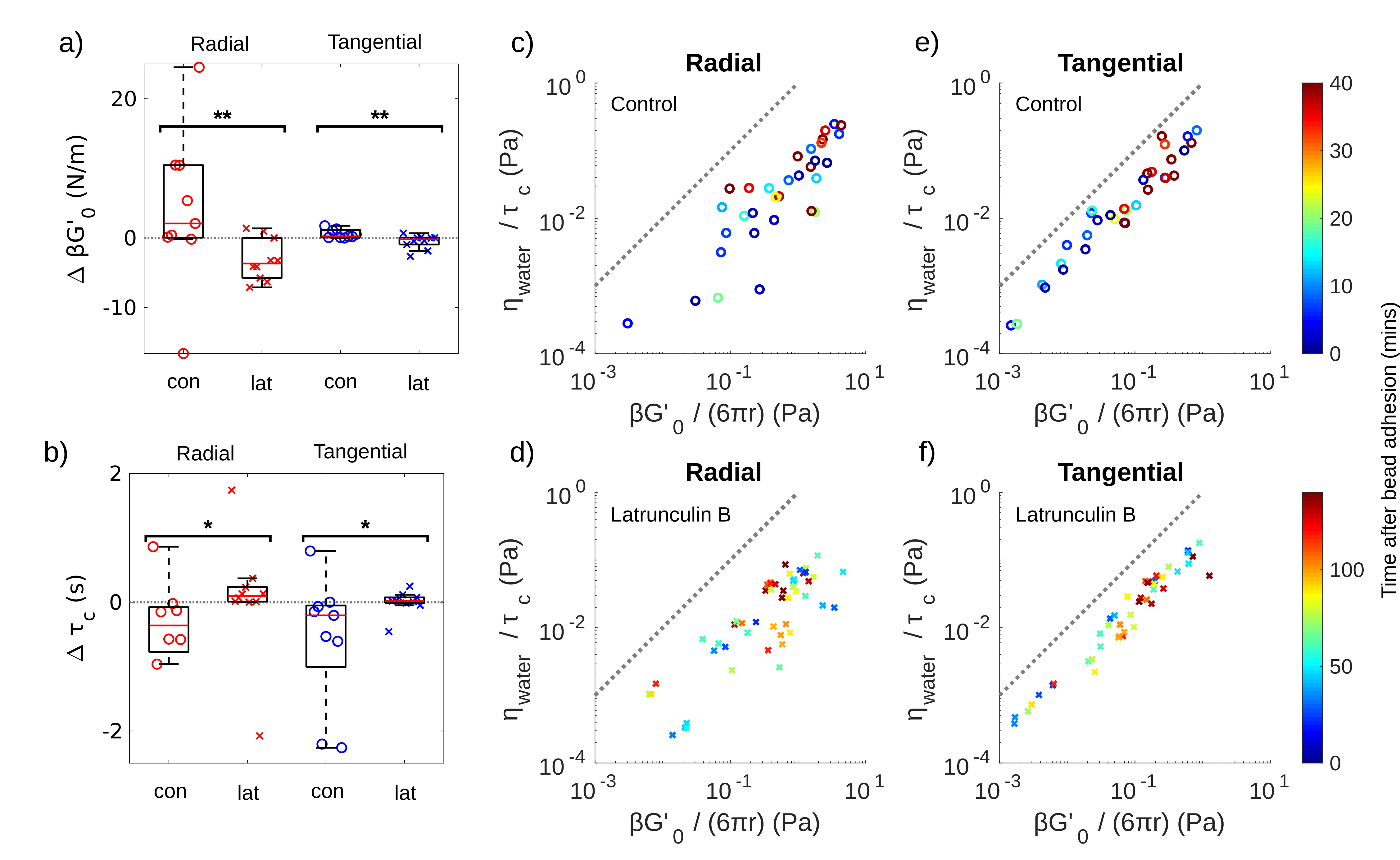}
    \caption{Master curves demonstrating constant viscosity across all cell measurements. (a,b) Change in geometric stiffness (eq. \ref{eq:cell_stiff}) and plateau onset time (eq. \ref{eq:trap_corner}), respectively, for control (con, circles) vs drug treated (lat, crosses) in radial (red) and tangential (blue) directions, respectively. ** denotes p $<$ 0.01 and * denotes p $<$ 0.05 with Mann-Whitney u-test. (c-f) Master curves of plateau onset time scaled by viscosity of water against geometric stiffness scaled by bead radius, for control (c,e) and drug treated (d,f). Dashed lines are a guide showing prediction for an optically trapped bead in water. Note that the time range of the colour scale for the control condition is different from that of the drug treated.}
    \label{fig:master_curves}
\end{figure}

\noindent
Consider now the viscoelastic shear properties of cells at timescales up to 1 second. The change over 40 minutes of the geometric stiffness, $\beta G'_0$, of the bead-cell system (eq. \ref{eq:cell_stiff}, figure \ref{fig:flowchart} box n) is shown in figure \ref{fig:master_curves}a, with the two perpendicular directions of motion analysed separately (radial and tangential directions identified by the red and blue symbols, respectively).
From figure \ref{fig:master_curves}a it can be seen that during the 40 minute experimental window the control cells stiffened, while the Latrunculin B treated cells softened. The tangential direction exhibited much smaller changes than the radial direction but showed the same trend. The distribution of stiffness changes for control and drug treated were found to be significantly different using a Mann-Whitney u-test (p $< 0.01$). Notably, in analogy to OT measurements, where an increase in trap stiffness causes a decrease in plateau onset time (eq. \ref{eq:trap_corner}, figure \ref{fig:flowchart} box k), the same behaviour is observed in the case of cells confirming that the control cells are stiffening over time whereas the drug treated cells are softening, as shown in figure \ref{fig:master_curves}b, with statistical significance (p $< 0.05$).

This observation is further corroborated by the results shown in figure \ref{fig:master_curves}c-f, where data is pooled together into two categories: ``no drug'' (con) and ``drug'' (lat) (no drug: 33 measurements from 23 cells; drug: 50 measurements from 10 cells), and the inverse of the corner time is plotted against the geometric stiffness of the cells. The ordinate axis has been multiplied by the viscosity value of water ($\eta_{water}=1$mPa s) and the abscissa axis has been scaled by a geometrical factor related to the bead radius ($6\pi r$), so that both the axis had the same dimensions and better represent equation \ref{eq:trap_corner}.
A visual inspection of figure \ref{fig:master_curves}c-f reveals that all cells have roughly equal and time-invariant viscosity as all measurements lay (over two decades) close to a single line of gradient 1 in a double logarithm plot (grey dotted line).

Notice that the colour gradient of the markers in figure \ref{fig:master_curves}c-f relates to the time passed after the bead adhesion, and in the case of the results shown in figure \ref{fig:master_curves}c-e highlights the stiffening over the time of cells without drug treatment. 
Moreover, by fitting the data shown in figure \ref{fig:master_curves}c-e to
\begin{equation}
    \log(\beta G'_0/6\pi r) = \log(\eta_{water}/\tau_c) + \log(\eta_r),
    \label{eq:etacell}
\end{equation}
we quantify the viscosity probed independently of changes in power-law rheology or elastic stiffness.
We found this viscosity for all cells in the radial direction to be $23\times\eta_{water}$ and in the tangential direction to be $4\times\eta_{water}$. Furthermore, we found that Latrunculin B treatment does not affect the viscosity probed.

\subsection{Non-equilibrium behaviour}
\noindent

\begin{figure}[t]
    \centering
    \includegraphics[width=0.9\textwidth]{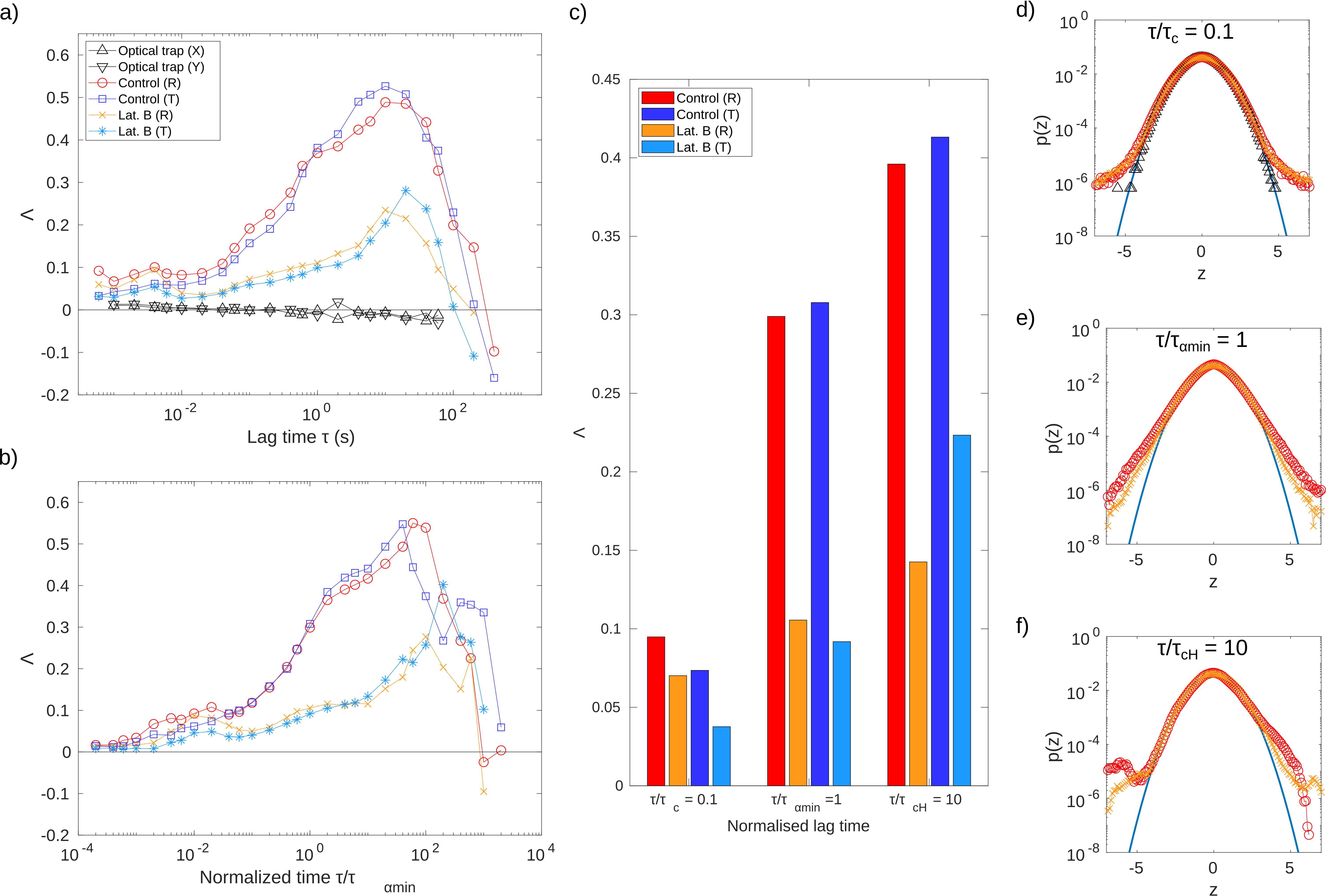}
    \caption{a) Non-gaussian parameter $\Lambda(\tau)$ as a function of lag-time, $\tau$, and b) as a function of normalised lag-time, $\tau / \tau_{\alpha min}$, where $\tau_{\alpha min}$ is the lag time where the MSD exhibits a power-law exponent minima. c) Non-gaussian parameter $\Lambda(\tau)$ at three different timescales as described in the text. Increment distributions for each time scale are shown as d-f).}
    \label{fig:ddist_and_ngp}
\end{figure}

The increment distribution and non-Gaussian parameter (equation \ref{eq:ddist}) were calculated and data pooled into control and latrunculin B treated (actin-disrupted). Temporal normalisations were performed after calculating increment distributions but before data pooling. In figure \ref{fig:ddist_and_ngp}a) we report the non-Gaussian parameter for each of the above cases in both radial and tangential directions, along with the case of an optically trapped bead for comparison. It can be seen that the increment distributions are increasingly broad-tailed at long lag-times (larger value of non-Gaussian parameter) and more so for the control case than actin-disrupted. 

In order to better understand the three characteristic regions of the MSD curve, three different characteristic times were chosen for consideration: (i) one-tenth of the plateau onset time, $\tau_c$, is chosen as representative of short time dynamics (figure \ref{fig:ddist_and_ngp}d)); (ii) the time coincident with the power-law exponent minimum as representative of the glassy rheology of the cytoskeleton (figure \ref{fig:ddist_and_ngp}e)); and (iii) ten times the superdiffuse onset time, $\tau_{cH}$, as representative of the superdiffuse motion (figure \ref{fig:ddist_and_ngp}f).

Normalisation of the delay time before accumulating the increment distributions allows characteristic times to be aligned, and the mean increment distributions to be inspected relative to these characteristic times. In figure \ref{fig:ddist_and_ngp}c) it can be seen that the broad-tailed increment distribution characteristic of non-equilibrium behaviour is coincident with the glassy plateau in the cell rheology (as seen on the MSD curve). While at shorter times, the bead motion is closer to equilibrium, with low values of $\Lambda(\tau)$ for all $\tau / \tau_{\alpha min} < 1$, thus indicating that the bead motion is driven by a single Gaussian process.

Examination of the increment distributions at short lag-times reveal a distribution close to Gaussian, with little difference between control and actin-disrupted cells. For both, the increment distribution departs from a normal distribution at circa $|z|=4$. As the delay time increases to the plateau time, the increment distribution from control cells develops broad tails, while the actin-disrupted cells exhibit a smaller increase. This is likely due to the reduced contribution of active processes within the disrupted cytoskeleton of the drug treated cells. At both the plateau time and longer timescales, the break from normal distribution occurs around $|z|=3$, indicating a systematic increase in large displacements relative to the shortest timescale. 

\begin{figure}
    \centering
    \includegraphics[width=0.9\textwidth]{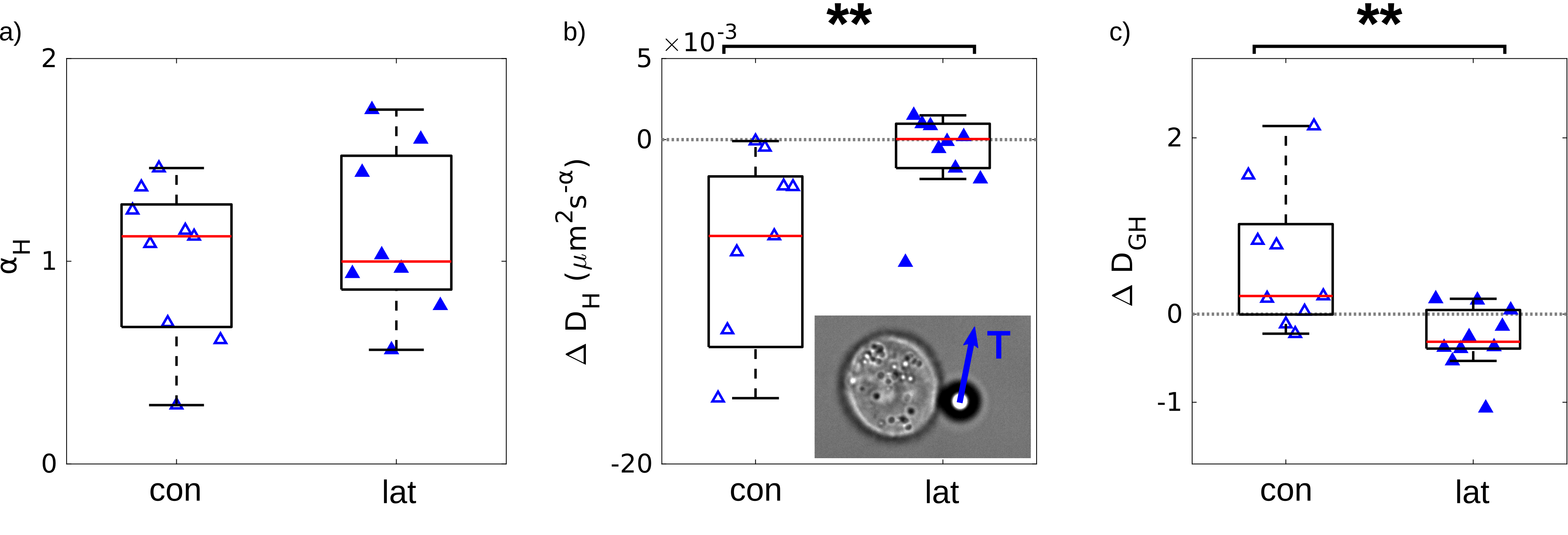}
    \caption{Pseudo-diffusion parameters $\alpha_H$ and $D_H$ (see equation \ref{eq:leastSq}) for long time dynamics in tangential direction, equivalent to the strain rate during cytoskeletal reorganisation. a) Long time motion of beads in the tangential direction tends to be superdiffuse, i.e. $\alpha_H > 1$. b) Bead motion decreases in time after bead adhesion for control case (con, empty triangles), and increases after drug treatment (lat, filled triangles). c) Normalisation removes the effect of changing cell stiffness, showing that the rate of cytoskeletal reorganisation increases in the control case, but that the bead motion decreases due to the increased cell stiffness. The reverse is true for the drug case. Statistically significant changes between drug and control are marked with ** for p $<0.01$ with Mann-Whitney u-test. Inset: Cell with bead and tangential direction labelled.}
    \label{fig:superdiff_params}
\end{figure}

To further characterise the non-equilibrium component of the bead motion, the long-time superdiffuse motion labelled region iii) in figure \ref{fig:cell_w_bead}e was analysed with least squares fitting (equation \ref{eq:leastSq}). The pseudo-diffusion coefficients are shown in figure \ref{fig:superdiff_params}. Inspection of the power-law exponent reveals that the motion over timescales of tens of seconds tends to be super-diffuse, i.e. $\alpha_H > 1$. This indicates correlated motion on long timescales, which would not be possible if the system were at thermal equilibrium. Previous passive microrheology studies of cells have attributed this superdiffusion to cytoskeletal reorganisation\cite{bursac2005cytoskeletal}.

Changes to long-time pseudo-diffusion parameters show a significant decrease in $D_H$ for the control cells, which means the internal force generation causes a lower strain rate within the cell. The reverse is true for the actin-disrupted cells, with a greater strain rate after drug treatment. However, this is not the whole picture. When the MSD is normalised by the geometric stiffness of the cell, the change in normalised pseudo-diffusion coefficient, $D_{GH}$, shows the opposite trend. In other words, the strain rate of control cells decreases because of the increase in the stiffness of the cell, and not because of a decrease in cytoskeletal rearrangement. Conversely in the actin-disrupted cells, the strain rate increases because the cell is softer while the cytoskeletal activity is reduced. This is indicative of the activity of molecular motors increasing with time after bead adhesion in the control case, while the drug-treated cells do not have the same ability to coordinate their cytoskeletal activity due to the disruptive action of the drug.

\section{Discussion}

\noindent In this work, we have demonstrated the effectiveness of a non-invasive passive microrheology technique for probing different processes at different timescales of living cells. We have presented a quantitative analysis of the time-dependent bead MSD and increment distribution based on an analogy to the well-established analytical framework used for optical trapping microrheology. Our approach enables quantitative analysis of cellular interactions with a functionalised surface, and the tracking of a cell's viscosity and elasticity over several hours.

We have observed cells stiffening after bead adhesion and softening after treatment with an actin depolymerising drug, while the viscosity probed remained invariant. The beads used were functionalised with streptavidin, a compound known to bind to both biotin and to integrin receptors\cite{alon1993cell,jurchenko2014integrin}. Cellular biochemistry literature\cite{wiesner2005integrin} reports that integrin binding by ligands triggers cytoskeletal reorganisation and connection to the actin cytoskeleton, and that disruption of actin prevents focal adhesion maturation. 

Focal complex formation, including actin recruitment, after integrin binding offers one plausible mechanism by which the binding of a bead could cause cells to stiffen\cite{wiesner2005integrin}. This correlates with the observed increase of the geometric stiffness after the binding of a bead to the cell. The softening of the cell observed after addition of Latrunculin B thus represents a partial loss of the contribution of actin filaments to the stiffness of the cell. No previous study, to our knowledge, has reported changes in mechanical properties during the maturation of focal adhesions.

Interpretation of results through the framework of passive microrheology assumes that the cell-bead system is in thermal equilibrium, meaning the MSD is equivalent to the shear creep compliance according to the fluctuation-dissipation theorem. Living cells are clearly far from equilibrium, however previous studies of the microrheology of active actin-myosin gels \cite{mizuno2008active} and of living cells\cite{lau2003microrheology} demonstrate that for time scales less than 1s (i.e.: regions i and ii on figure \ref{fig:cell_w_bead}e), thermally excited motion dominates and the generalised Stokes-Einstein relation (equation \ref{eq:LangvinWarrenSol2}) is still applicable. This is reflected in the increment distributions which are near-Gaussian for short lag times and broad tailed for longer lag times.

The mechanical properties of the cytoskeleton can be measured using active microrheology with extracellular probes, giving rise to the soft glassy rheology model\cite{Fabry2001}. Similar power-law scaling was found using passive microrheology by tracking endogenous intracellular probes\cite{lau2003microrheology}. These results validate the interpretation of our passive microrheology using extracellular probes to test the mechanical properties of the cytoskeleton.

Prior passive microrheology with extracellular probes have suggested cytoskeletal rearrangement as the cause of superdiffuse motion at long times (MSD power-law exponent $\alpha_H>1$), as also reported in this work. However, studies did not report measurements of displacements over timescales less than a second\cite{an2004role,bursac2007cytoskeleton,lenormand2011dynamics}. An optical trap can be used to control bead binding to cells, allowing us probe the mechanical properties of the binding site at time scales of less than a millisecond, limited by the frame rate of the camera, while tighter binding of the bead to the cell would result in motion being below the detection limit on millisecond timescales. Passive microrheology can to track the mechanical properties of the cytoskeleton unperturbed by rejuvenation effects caused by deformation under applied forces\cite{bursac2005cytoskeletal,trepat2007universal}.

The viscosity inferred from equation \ref{eq:trap_corner} was found to be constant for all measurements. Invariant viscosity of integrin complexes has been reported before using an active microrheology method\cite{bausch1998local}, but this is the first time it has been reported using a passive method. We also observe that although disrupting the actin cytoskeleton with Latrunculin B causes a significant reduction in the geometric stiffness, there is still no significant change to the viscosity. Thus the viscosity probed is independent of cytoskeletal changes induced in our experiments by the addition of Latrunculin B. 

At time-scales around a second, the cell behaves as a soft glassy material exhibiting weak power-law scaling; at these time-scales the microrheology may only probe longitudinal modes of the cytoskeleton rather than shear modes of the cytoskeleton and cytosol\cite{levine2000one}. This can be understood by considering the behaviour of the cytosol when the cytoskeleton is displaced through it: at sufficiently low frequencies, the cytosol is able to drain freely, and its viscosity will no longer affect deformations of the cytoskeleton.

At longer time-scales, the microrheology probes the active strain fluctuations of the cell\cite{lau2003microrheology,an2004role}, which are a result of cytoskeletal rearrangements \cite{trepat2007universal,bursac2007cytoskeleton}. Thus the MSDs at lag-times of greater than $10$s reported here are higher than if all motion were thermally excited. These slow strain fluctuations are shown to decrease in magnitude in parallel to the stiffening of the cell, while the power law exponent decreases in time after bead adhesion in almost all cases.

\section{Conclusion}
\noindent
In this work we have presented a quantitative analysis of the changes in mechanical properties of living cells by means of non-invasive passive microrheology measurements. These were performed for up to two hours after a microsphere was bound to the outside of a cell. This is, to our knowledge, the first time that a microrheological technique has been used to observe cytoskeletal changes over such extended time scale, and the first report of cell stiffness changes induced by integrin binding. 

The pseudo-Brownian motion of the probe particle was monitored over an experimental time window ranging from $10^{-3}$ to $10^{3}$ seconds and the MSD was interpreted through an analytical framework based on a Generalised Langevin Equation, which has been validated in literature in the case of microrheology with optical tweezers. The approach used in this work allows us to probe the high frequency mechanical properties of cells without the need of external forces. Non-drug treated control cells were seen to become stiffer over tens of minutes, which is interpreted as a result of the cytoskeletal reorganisation induced by integrin binding. This stiffening is prevented by treating the cells with Latrunculin B, an actin depolymeriser. This interpretation is supported by literature on streptavidin-integrin binding\cite{alon1993cell}, and the interplay between integrins and the actin cytoskeleton\cite{wiesner2005integrin}. 

Notably, our results are in agreement with previous microrheology studies of live cells. In particular, we report (i) constant viscosity during normal cell functions, (ii) a weak power-law scaling in the elastic modulus, and (iii) active fluctuations at long time scales. In addition we identify time-scales where the thermal fluctuations of the cell dominates and those were the dynamics of the cell are governed by actively (driven) processes, how these regime change are interpreted and discussed. 


\section{Materials and methods}

\small
\noindent Hela S3 (ATCC CCL-2) cells were grown in Dulbecco's Modified Eagle Medium, supplemented with 10\% (v/v) Fetal Bovine Serum (Merck: F9665), 1\% (v/v) L-glutamine (Merck: G7513), 1\% (v/v) Penicillin-Streptomycin solution (Merck: P0781) and 1\% (v/v) Non-essential Amino Acid Solution (Gibco:11140-050); these were grown in a T75 flask using the same protocols as for adherent cells. A portion of the cells were adhered to the flask at any time, and these cells were used for experiments while the cells which remained in suspension were used to continue the cell line.

The Hela S3 cell line was chosen as they adhere to glass coverslips but remain rounded (as shown in figure \ref{fig:confocal}, allowing the bead to be placed where the cell surface is normal to the imaging plane, shown in figure \ref{fig:cell_w_bead}. Adhesion to the coverslip is essential to prevent Brownian motion of the cell during the experiment.

Confocal microscopy (figure \ref{fig:confocal}) was performed on a Zeiss LSM900 microscope with Airyscan 2 detector, using a 20x 0.6NA objective lens. Cells were seeded 24 hours in advance at a density of $10^5$ cells per 25mm coverslip coated with PLL (Merck, P04707) as described by the manufacturer to promote cell adhesion, fixed in 4\% formaldehyde solution (Merck, 1.00496), permeabilised with 0.1\% Triton X-100 and stained with Rhodamine Phalloidin (Thermofisher Scientific R415) and Hoescht 33342 (Thermofisher Scientific, H3570) for actin and DNA respectively. This method was based on a protocol by Mitchison et al\cite{MitchisonLabWebsite}. To image the effects of Latrunculin B (Merck, L5288), the cells were incubated with the drug at 1mM in serum-free DMEM for 10 minutes before fixing.

Before microrheology experiments, $10^4$ cells per well were seeded into an 8 well cavity microslide (Ibidi, Germany, 80841) with 500 \textmu L of growth media as above. Cells were incubated overnight at 37$\degree$C with 5\% CO$_2$, and 2 hours before experiments the media was changed for 200 \textmu L of serum-free Minimum Essential Medium Eagle (Merck: M2279) supplemented as for the growth media (with the exception of FBS) and 5$\mu$g Calcein-AM (ThermoFisher, USA, C3099) for fluorescent viability imaging.

Optical trapping and microrheology experiments were performed on a custom microscope set up as follows: a 3W 1064nm ND:YAG was coupled into the back focal plane of an inverted microscope (Nikon, Japan). A 1.3NA, 100x oil immersion objective (Nikon, Japan) is used to focus the beam into the sample, which was mounted on a motorised XY stage (MS-2000, ASI Instruments). The optical trap was used to position polystyrene beads (Spherotech, USA) relative to individual cells. Imaging was performed in transmission using the same objective to form an image on an sCMOS camera (QImaging, Canada). 

The image acquisition was controlled using $\mu$Manager\cite{Edelstein2010}. Data collection for each microrheology measurement was as follows: first, a full field of view image is collected, before starting a sequence acquisition with reduced field of view (see figure \ref{fig:cell_w_bead}a iv for example), allowing frame rates over 1,000 frames per second. To find the bead position in each image, first the background is removed by subtracting a value (manually chosen before the acquisition) from every pixel (pixel values below the threshold become 0, not negative), and the image centroid is calculated in Java using code adapted from the ImageJ source code\cite{schneider2012nih}.

A typical video measurement comprises of 1 million images over around 9 minutes. The centroid co-ordinates from every image are stored during acquisition, for example figure \ref{fig:cell_w_bead}c. The image and metadata for only every thousandth image are retained for future reference thus reducing storage and memory footprint. After the sequence acquisition, the field of view on the camera is reset and a second full field of view image is collected. Finally, data is saved to disk for later analysis. 

\section{Acknowledgements}
This work was supported by the Engineering and Physical Sciences Research Council (EPSRC) under grants EP/S0214334/1, \\
EP/L016052/1, EP/R035067/1, EP/R035563/1, EP/R035156/1 "Experiencing the micro-world: a cell's perspective", the Oxford and Nottingham Biomedical Imaging (ONBI) CDT and the University of Nottingham. The authors thank the Nanoscale and Microscale Research Centre (nmRC) for providing access to instrumentation for confocal microscopy experiments.

\bibliographystyle{MSP}
\bibliography{refs}

\end{document}